\documentclass[twocolumn]{aastex631}

\newcommand{\hi}{H\,{\sc i}}
\newcommand{\kms}{km\,s$^{-1}$}
\newcommand{\jykms}{Jy\,km\,s$^{-1}$}
\newcommand{\jybmkms}{Jy\,beam$^{-1}$km\,s$^{-1}$}

\received{June 17, 2025}
\accepted{January 26, 2026}

\begin{document}

\title{High resolution observations of `dark' neutral hydrogen clouds in the Virgo cluster with the Very Large Array\footnote{The National Radio Astronomy Observatory is a facility of the National Science Foundation operated under cooperative agreement by Associated Universities, Inc.}}

\author[0000-0002-1261-6641]{Robert Minchin}
\affiliation{National Radio Astronomy Observatory, 1011 Lopezville Rd.,
P.O. Box O, Socorro, NM 87801, USA}
\author[0000-0002-3782-1457]{Rhys Taylor}
\affiliation{Astronomical Institute of the Czech Academy of Sciences, Bo\v{c}n\'{i} II 1401/1a, 141 00 Praha 4, Czech Republic}
\author[0000-0003-3168-5922]{Emmanuel Momjian}
\affiliation{National Radio Astronomy Observatory, 1011 Lopezville Rd.,
P.O. Box O, Socorro, NM 87801, USA}
\author[0000-0002-7898-5490]{Boris Deshev}
\affiliation{Tartu Observatory, Faculty of Science and Technology, University of Tartu, Estonia}
\author[0000-0002-2294-9263]{Vojt\v{e}ch Part\'{i}k}
\affiliation{Astronomical Institute of the Czech Academy of Sciences, Bo\v{c}n\'{i} II 1401/1a, 141 00 Praha 4, Czech Republic}
\affiliation{Astronomical Institute of Charles University, V Hole\v{s}ovi\v{c}k\'{a}ch 747/2, 180 00 Praha 8, Czech Republic}

\correspondingauthor{Robert Minchin}\email{rminchin@nrao.edu}



\begin{abstract}

We have observed six `dark' neutral hydrogen (\hi) clouds discovered in the Virgo cluster by the Arecibo Galaxy Environment Survey (AGES) with the Karl G. Jansky Very Large Array (VLA), giving higher angular and velocity resolution than the original AGES observations. We detected compact \hi\ emission in two of the sources, AGESVC1~231 and AGESVC1~274, allowing us to firmly associate them with faint ($m_g > 18.5$), blue ($g-i < 0.1$) optical counterparts with high $M_{HI}/L_g$ ratios. In a further two sources, we detected low column-density extended \hi\ emission, consistent with these being dispersing clouds from ram-pressure stripping or tidal interactions. The final two sources were not detected with the VLA, allowing us to set low column-density limits on the \hi\ detected by AGES that are consistent with these clouds also being formed from \hi\ that is dispersing into the intra-cluster medium. The four \hi\ sources not associated with optical counterparts thus appear likely to be relatively short-lived objects. No evidence was found for either pressure-supported turbulent spheres or stable dark galaxies.

\end{abstract}

\keywords{}


\section{Introduction} \label{sec:intro}

The Arecibo Galaxy Environment Survey (AGES) was a deep neutral hydrogen (\hi) survey carried out with the 305-m William E. Gordon Telescope at Arecibo Observatory using the Arecibo L-band Feed Array (ALFA) multibeam system, which covered 200 square degrees in multiple regions on the sky. This included 25 square degrees in the Virgo Cluster \citep{2012MNRAS.423..787T,2013MNRAS.428..459T}, where it reached a $3\sigma$ column-density limit of $4.5 \times 10^{17} {\rm cm}^{-2} \left(\frac{\Delta V}{\rm 10\ km\,s^{-1}}\right)^{0.5}$.

AGES discovered a number of sources in the Virgo Cluster that did not have obvious optical counterparts and were thus identified as potential optically dark clouds. The Virgo Cluster has been a rich source of optically dark and almost dark clouds \citep[e.g.,][]{1989ApJ...346L...5G,2005ApJ...622L..21M,2007ApJ...665L..15K}. Many of these have later been identified with optical counterparts \citep[e.g.,][]{1990ApJ...351L..33I,2024ApJ...966L..15J} or explained as tidal debris \citep[e.g.,][]{2008ApJ...673..787D}, but the high velocity widths of some of the AGES clouds and their lack of connection (at the AGES column-density limit) to optically bright galaxies has made them particularly challenging to explain. 

Models proposed for the origin of these sources have included fragmentation of long \hi\ streams formed by tidal interactions, (almost) dark galaxies, and pressure-supported turbulent spheres \citep{2016MNRAS.461.3001T,2017MNRAS.467.3648T,2018MNRAS.479..377T,2016ApJ...824L...7B}. However, simulations of fragmentation of long \hi\ streams by harassment have shown that this produces a negligible fraction of clouds with high velocity widths, and those produced were short lived with detectable lifespans in the tens of megayears, while simulations of turbulent spheres similarly found that these were unstable, with detectable lifespans under $\sim 100$ megayears. Both of these explanations would thus imply either a continually renewing population of such clouds or that we are observing a special time in the Virgo Cluster.

Here, we present the results of observations of six of these \hi\ sources carried out with the NSF's Karl G. Jansky Very Large Array (VLA). These observations were planned to have an angular resolution of $\sim$ 50\arcsec\ in order to allow us to identify faint 
optical counterparts to almost dark galaxies or to see distinctive structures in the \hi\ distribution that could give clues as to their origin in fragmented streams or turbulent spheres, while retaining sensitivity to low column-density gas.

\section{Observations} \label{sec:obs}

Observations were carried out with the VLA in D configuration in February 2017, using the WIDAR correlator with 8-bit sampling and 4096 channels over a 32~MHz bandwidth, giving a channel separation of 7.8125~kHz (1.7~\kms). The flux density scale calibrator used was 3C286, and the complex gain calibrator was J1254+1104. Data reduction was carried out in CASA \citep{2022PASP..134k4501C}, including editing out bad data, calibration, and continuum subtraction at the positions of M87 and M49, as well as at the position of each target source. 

Different settings in CASA's {\sc tclean} task were used depending on the specific requirements of each source to make the images. For the two sources that showed compact \hi\ emission, which were the two with high enough signal-to-noise to allow {\sc clean}ing, Briggs weighting was used with {\sc robust = 0.5} and the cube was {\sc clean}ed within a user-defined region. For the source AGESVC1~231, four channels were binned together for this {\sc clean}ing in order to enhance the signal-to-noise ratio. For the four sources that showed either low brightness temperature extended \hi\ emission or no detectable \hi\ emission, Briggs weighting was used with {\sc robust = 0.8} (giving a good balance between signal to noise and side-lobe suppression), a UV taper was set to give a beam size of $\sim 100$\arcsec\ (i.e. half of the Arecibo beamsize of $\sim 200$\arcsec\ in the AGES cubes), and no {\sc clean}ing was performed, considering the low signal-to-noise ratios in these sources. A 5 percent flux density scale calibration accuracy was assumed for the VLA.

The noise level reached (prior to applying any UV taper) was 2.0--2.3~mJy per channel for all sources except for AGESVC1~262, where the noise level was 3~mJy. This is consistent with the on-source time of around 1.75h per source and robust weighting. Untapered robust beams were a little larger than $50\arcsec \times 40\arcsec$, with some variation between sources. This sensitivity and beam size gives a $3\sigma$ column-density of approximately $1.5\times 10^{19} {\rm cm}^{-2} \left(\frac{\Delta V}{\rm 10\ km\,s^{-1}}\right)^{0.5}$ (50 percent higher than this for AGESVC1~262) in the untapered cubes, around 30 times higher than AGES but still an order of magnitude below the column densities seen within the optical disks of galaxies. As noted above, application of UV tapers was used to increase the column-density sensitivity in the image cubes showing extended emission or no detectable emission.

\section{Results} \label{sec:results}
Of the six sources, four were either firm or probable detections and results for these are presented here. We remeasured the properties of all of the sources in the AGES data cube, which were originally measured in a `production line' model as part of the survey; this enabled us to spend more time refining the profile window and thus getting more accurate measurements of the line widths and total \hi\ fluxes. Where there was a clear central peak in velocity with extended wings at lower flux levels, as in the cases of AGESVC1~231, AGESVC1~258 and AGESVC1~266, the value chosen for $\Delta V_{50}$ was the width across the profile at the first point where the flux fell below 50\% of the peak rather than across the profile at the first point where the flux rose to 50\% of the peak moving in from the edges of the profile window. This was done in order to be conservative in measuring the velocity width where, on relatively low signal-to-noise spectra, the velocity width might have been boosted by the inclusion of noise spikes on the extended lower-flux-level wings. For all but AGESVC1~231, the remeasurement followed the standard AGES procedure of using {\sc mbspect} within {\sc miriad} with ``yaxis = point'' and ``width = 5'' to recreate the flux of a point source (to Arecibo) within a $5\times 5$ pixel (i.e., $5\arcmin\times 5\arcmin$, as the AGES pixels are 1\arcmin\ in size) region. AGESVC1~231 was found to be extended in the AGES data when this was re-examined and so was analyzed as an extended source, as described below. The original AGES measurements and the results of our remeasurements are given in table \ref{agestable}, along with the assumed distance to each source \citep[from][]{2012MNRAS.423..787T} and the calculated \hi\ masses from the original and remeasured fluxes. As described in \citet{2012MNRAS.423..787T}, section 4.2, the assumed distance depends on the sub-cluster to which each source is assigned, at distances of 17 (the main cluster body), 23 or 32 Mpc; as there is uncertainty in these assignments, there is a corresponding uncertainty in the distances. For our analysis, we assume the assignments of sources to sub-clusters, and thus the distances to those sources, in \citet{2012MNRAS.423..787T} to be correct and adopt them here.

For the two sources not detected with the VLA, we used moment 0 maps formed from un-{\sc clean}ed VLA data at a resolution of $\sim100$ arcsecond over the 50 percent velocity width of the remeasured AGES data to calculate $3\sigma$ column-density thresholds. Table \ref{resultstable} gives the the position of the VLA \hi\ peak (if detected), the assumed distance to the source, the VLA \hi\ fluxes or upper limits calculated from the $3\sigma$ column-density limits and assuming a source size the same as the AGES beam size for the non-detections, the peak column-density (averaged over the beam, so a lower limit for unresolved sources) or upper limit, the VLA \hi\ mass or upper limit calculated from the VLA \hi\ flux, and the velocity widths measured at 50 percent and 20 percent of the peak flux density in the \hi\ spectrum.

\begin{deluxetable*}{lccccccccc}
\caption{Summary of the original and remeasured AGES data\label{agestable}}
\tablehead{
\colhead{Source}&\multicolumn{5}{c}{Original}&\multicolumn{4}{c}{Remeasured}\\
&\colhead{$F_{HI}$}  &\colhead{$M_{HI}$}&\colhead{$\Delta V_{50}$}&\colhead{$\Delta V_{20}$}&\colhead{D}&\colhead{$F_{HI}$}  &\colhead{$M_{HI}$}&\colhead{$\Delta V_{50}$}&\colhead{$\Delta V_{20}$}\\
&\colhead{\jykms}&\colhead{$10^7~\nom{M}$}&\colhead{\kms}&\colhead{\kms}&\colhead{Mpc}&\colhead{\jykms}&\colhead{$10^7~\nom{M}$}&\colhead{\kms}&\colhead{\kms}
}
\startdata
AGESVC1~231&$0.17\pm0.05$&$4.2\pm1.2$&$\phn36\pm20$&$152\pm30$&32&$0.51\pm0.06$&$12.3\pm1.5$&$35\pm\phn6$&$189\pm21$\\
AGESVC1~257&$0.20\pm0.06$&$1.3\pm0.4$&$131\pm13$&$157\pm30$&17&$0.21\pm0.03$&$\phn1.4\pm0.2$&$71\pm12$&$\phn97\pm17$\\
AGESVC1~258&$0.20\pm0.05$&$1.3\pm0.4$&$\phn32\pm18$&$120\pm27$&17&$0.27\pm0.04$&$\phn1.8\pm0.3$&$30\pm15$&$134\pm27$\\
AGESVC1~262&$0.16\pm0.05$&$2.0\pm0.7$&$104\pm13$&$146\pm19$&23&$0.18\pm0.04$&$\phn2.2\pm0.3$&$35\pm11$&$\phn77\pm17$\\
AGESVC1~266&$0.25\pm0.06$&$1.7\pm0.4$&$\phn77\pm21$&$173\pm32$&17&$0.31\pm0.05$&$\phn2.1\pm0.3$&$22\pm18$&$174\pm23$\\
AGESVC1~274&$0.11\pm0.03$&$0.7\pm0.2$&$\phn22\pm\phn4$&$\phn35\pm\phn6$&17&$0.18\pm0.02$&$\phn1.2\pm0.2$&$22\pm\phn4$&$\phn35\pm\phn6$
\enddata
\end{deluxetable*}

\begin{deluxetable*}{lcccccccc}
\caption{Summary of VLA observations\label{resultstable}}
\tablehead{
\colhead{Source}&\colhead{R.A.}&\colhead{Decl.} &\colhead{D}&\colhead{$F_{HI}$}&\colhead{$N_{HI}$ (peak)}&\colhead{$M_{HI}$}&\colhead{$\Delta V_{50}$}&\colhead{$\Delta V_{20}$}\\
&&&\colhead{Mpc}&\colhead{\jykms}&\colhead{$10^{18}$ cm$^{-2}$}&\colhead{$10^7~\nom{M}$}&\colhead{\kms}&\colhead{\kms}
}
\startdata
AGESVC1~231&$12^{\mathrm h}18^{\mathrm m}14^{\mathrm s}$&$07\arcdeg21\arcmin40\arcsec$&32&$0.20\pm0.03$&$75\pm15$&$4.8\pm0.7$&$36\pm13$&$60\pm20$\\
AGESVC1~257&\nodata&\nodata&17&$<0.44$&$<11$&$<3.0$&\nodata&\nodata\\
AGESVC1~258&$12^{\mathrm h}38^{\mathrm m}11^{\mathrm s}$&$07\arcdeg30\arcmin40\arcsec$&17&$0.29\pm0.05$&$15.6\pm3.6$&$2.0\pm0.3$&$74\pm\phn9$&$80\pm14$\\
AGESVC1~262&\nodata&\nodata&23&$<0.35$&$<8.9$&$<4.4$&\nodata&\nodata\\
AGESVC1~266&$12^{\mathrm h}26^{\mathrm m}12^{\mathrm s}$&$08\arcdeg00\arcmin50\arcsec$&17&$0.11\pm0.04$&$4.3\pm1.0$&$0.8\pm0.2$&$13\pm10$&$23\pm14$\\
AGESVC1~274&$12^{\mathrm h}30^{\mathrm m}36^{\mathrm s}$&$08\arcdeg38\arcmin40\arcsec$&17&$0.21\pm0.03$&$95\pm12$&$1.4\pm0.2$&$23\pm\phn6$&$39\pm\phn9$
\enddata
\end{deluxetable*}

\subsection{Optical counterparts}

Table \ref{opttable} gives the positions and the SDSS DR18 observed ({\it g}-band magnitude and observed $g-r$ and $g-i$ colors) for the optical counterparts identified in this study, while table \ref{optquanttable} gives the derived optical quantities. These include the extinction-corrected $g-i$ color, extinction-corrected {\it g}-band and {\it i}-band luminosities, the \hi\ mass to {\it g}-band luminosity ratio, the stellar mass, and the \hi\ to stellar mass ratio. Extinction corrections were made using the dust measurements of \citet{2011ApJ...737..103S}, via the IRSA Galactic Dust Reddening and Extinction service\footnote{\url{https://irsa.ipac.caltech.edu/applications/DUST/}}. Absolute luminosities were calculated using solar absolute magnitudes from \citet{2018ApJS..236...47W}. Stellar masses were calculated from the {\it i}-band luminosity and the $g-i$ color using equation \ref{mstar}, following \citet{2020AJ....160..271D} and the method of \citet{2011MNRAS.418.1587T}.

\begin{equation}
    \log(M_\star) = -0.68 + 0.70(g-i) + \log(L_i\label{mstar})
\end{equation}

However, the errors on the stellar mass and the \hi\ to stellar mass ratio are large, and the corrections described in \citet{2020AJ....160..271D} to remove systematic differences between the Taylor-method stellar mass estimates and those from the GALEX-SDSS-WISE Legacy Catalog 2 \citep{2016ApJS..227....2S} and from \citet{2015ApJ...802...18M} give results differing by more than a factor of two; a significant difference for AGESVC1~231. Direct calculation of the stellar mass of AGESVC1~231 from the WISE W1 luminosity following \citet{2015ApJ...802...18M} gives a result of $1.16 \pm 0.36 \times 10^7~\nom{M}$, close to that given by \citet{2020AJ....160..271D}'s conversion formula but more than twice that derived using the Taylor method.

\begin{figure*}
\plottwo{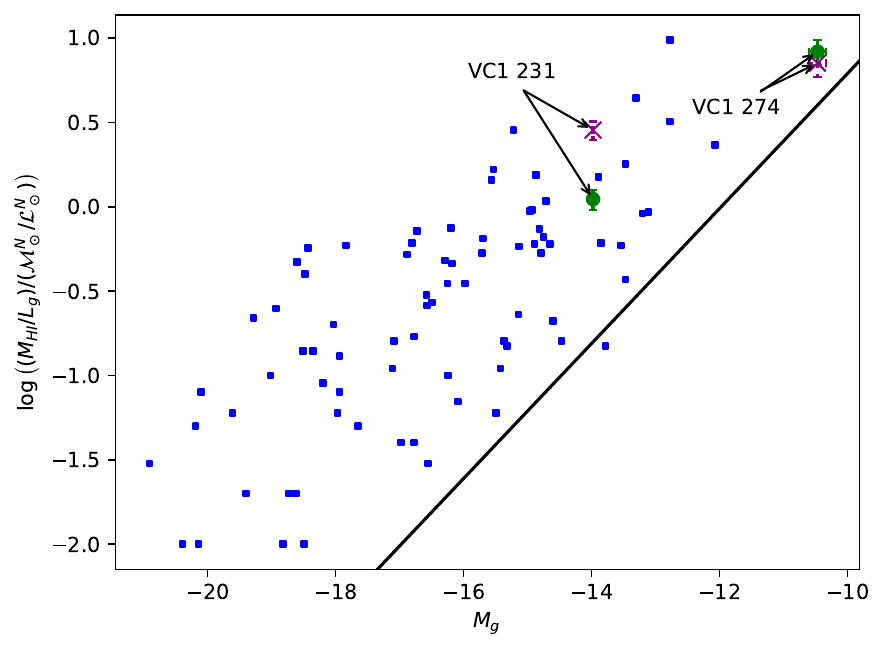}{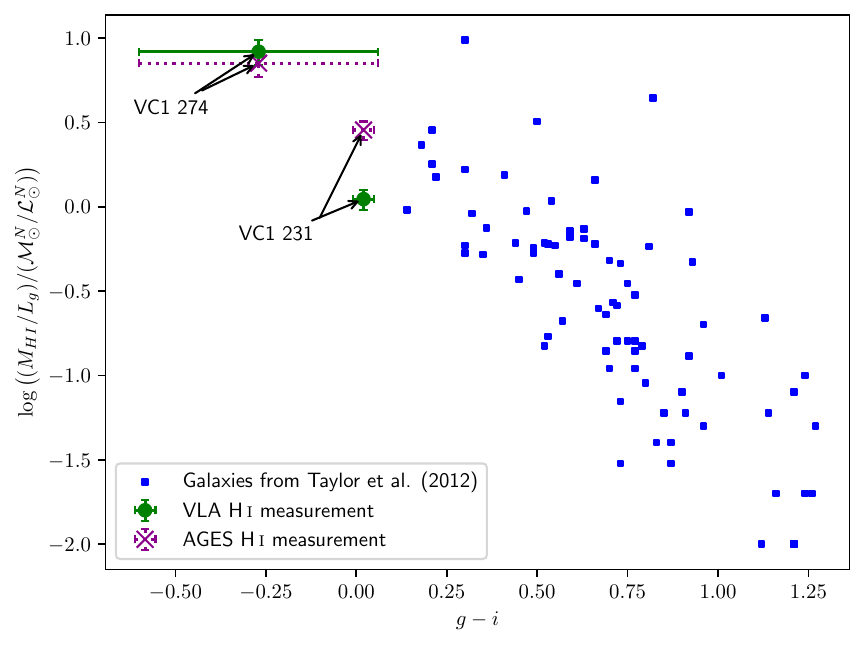}
\plottwo{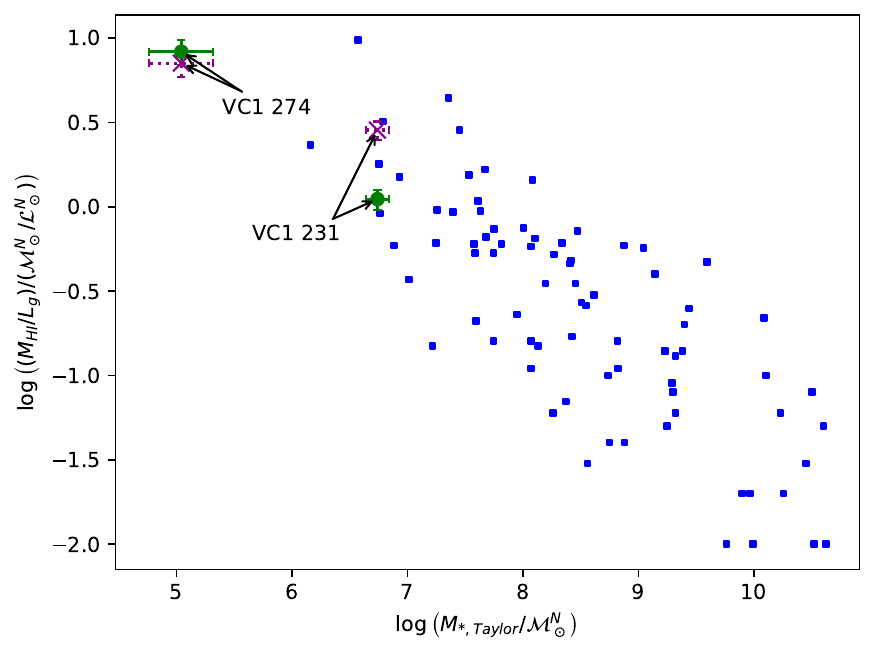}{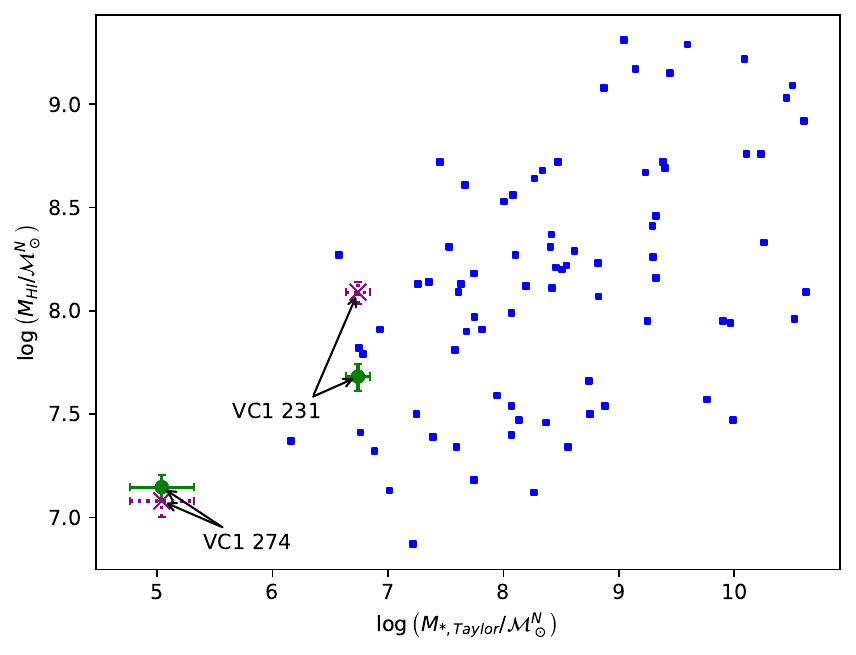}
    \caption{Plots of $M_{HI}/L_g$ versus $M_{g}$ (upper left; c.f. \citet{2012MNRAS.423..787T} figure 15), $g-i$ color (upper right; c.f. \citet{2012MNRAS.423..787T} figure 16), and $M_{\star\rm,Taylor}$ (lower left), and $M_{HI}$ versus $M_{\star\rm,Taylor}$ (lower right), comparing the AGES detections in the Virgo Cluster from \citet{2012MNRAS.423..787T} (blue squares) with the two sources with identified optical counterparts, for our VLA measurements (green circles with solid errorbars) and AGES remeasurements (magenta crosses with dotted errorbars) of the \hi\ content. The plot of $M_{HI}/L_g$ versus $M_{g}$ includes the approximate selection limit at 17~Mpc from \citet{2012MNRAS.423..787T} -- AGES was not sensitive to faint, gas-poor galaxies in the bottom-right of this plot and in the bottom-left of the plot of $M_{HI}/L_g$ versus $M_{\star\rm,Taylor}$.}
    \label{fig:mhilg_plots}
\end{figure*}

Figure \ref{fig:mhilg_plots} shows how the positions of these two objects (using the AGES and VLA measurements) compare to other AGES detections in the Virgo Cluster from \citet{2012MNRAS.423..787T}. It can be seen that they have high $M_{HI}/L_g$ ratios as well as low stellar masses and very blue colors. However, despite this they fall on the same relationships defined by the other galaxies from AGES -- their blue colors, faint optical magnitudes and low stellar masses are all in keeping with their high $M_{HI}/L_g$ ratios. In all cases, AGESVC1~274, the lower \hi\ mass galaxy, is the more extreme in terms of its optical properties.

Table \ref{uvirtable} gives the UV and IR properties of the counterparts from GALEX \citep{2017ApJS..230...24B} and AllWISE. Upper limits in the W2 band for AGESVC1~231 are 95\% confidence (i.e., $2\sigma$) upper limits from AllWISE profile-fitting photometry; AGESVC1~274 is not detected in AllWISE; upper limits given here are 95\% confidence limits from our own aperture photometry using the AllWISE standard 8\farcs5 radius. As a result, these limits are brighter than the limit and detection for AGESVC1~231.

\begin{deluxetable*}{lcccccl}
\caption{Observed optical properties\label{opttable}}
\tablehead{
\colhead{Source}&\colhead{R.A.}&\colhead{Decl.} &\colhead{$g$}&\colhead{$g-r$}&\colhead{$\phm{-}g-i$}&\colhead{ID}
}
\startdata
AGESVC1~231&$12^{\mathrm h}18^{\mathrm m}13\fs93$&$07\arcdeg21\arcmin47\farcs3$&$18.62\pm0.01$&$0.03\pm0.02$&$\phm{-}0.05\pm0.03$&SDSS~J121813.93+072147.3\\
AGESVC1~274&$12^{\mathrm h}30^{\mathrm m}26\fs40$&$08\arcdeg38\arcmin40\farcs2$&$20.73\pm0.11$&$0.42\pm0.16$&$-0.23\pm0.33$&SDSS~J123026.40+083840.2\\
\enddata
\end{deluxetable*}

\begin{deluxetable*}{lccccccc}
\caption{Derived optical properties\label{optquanttable}}
\tablehead{
\colhead{Source}&\colhead{D}&\colhead{$\phm{-}g-i$}&\colhead{$L_g$}&\colhead{$L_i$}&\colhead{$M_{HI}/L_g$}&\colhead{$M_{\star\rm,Taylor}$}&\colhead{$M_{HI}/M_{\star\rm,Taylor}$}\\
&\colhead{Mpc}&\colhead{$\phm{-}$(ext. corr.)}&\colhead{$10^7~\nom{L}$}&\colhead{$10^7~\nom{L}$}&\colhead{$\nom{M}/\nom{L}$}&\colhead{$10^6~\nom{M}$}
}
\startdata
AGESVC1~231~(VLA)&32&$\phm{-}0.02\pm0.03$&$4.31\pm0.07$&$2.57\pm0.05$&$1.11\pm0.15$&$5.5\phn\pm1.3\phn$&$\phn\phn8.6\pm\phn2.3$\\
\phm{AGESVC1~231}~(AGES)&&&&&$2.85\pm0.35$&&$\phn21.2\pm\phn5.7$\\
AGESVC1~274~(VLA)&17&$-0.27\pm0.33$&$0.17\pm0.02$&$0.08\pm0.02$&$8.3\phn\pm1.4\phn$&$0.11\pm0.07$&$132\phd\phn\pm88\phd\phn$\\
\phm{AGESVC1~274}~(AGES)&&&&&$7.1\phn\pm1.2\phn$&&$109\phd\phn\pm72\phd\phn$\\
\enddata
\end{deluxetable*}

\begin{deluxetable*}{lcccccc}
\caption{Observed ultraviolet and infrared properties\label{uvirtable}}
\tablehead{
\colhead{Source}&\colhead{GALEX ID}&\colhead{FUV mag}&\colhead{NUV mag}&\colhead{AllWISE ID}&\colhead{W1 mag}&\colhead{W2 mag}
}
\startdata
AGESVC1~231&J121813.9+072147&$19.13\pm0.14$&$19.42\pm0.07$&J121813.91+072146.6&$17.26\pm0.15$&$>16.787$\\
AGESVC1~274&J123026.5+083840&$21.68\pm0.44$&$21.88\pm0.39$&\nodata&$>16.393$&$>14.767$\\
\enddata
\end{deluxetable*}

\subsection{AGESVC1~231}
AGESVC1~231 is thought to be part of an infalling sub-cluster at 32 Mpc, behind the main Virgo cluster \citep{2012MNRAS.423..787T}. The original AGES measurements found an \hi\ flux of $0.173 \pm 0.050$~\jykms\ and a velocity width at 20\% of the peak of $152\pm30$~\kms.

This source was detected with the VLA and a clean map made, averaging 4 channels together to give a velocity resolution of 6.7~\kms, from which a moment 0 map was formed over 1875 to 1945~\kms. The restoring beam was $54\arcsec\times41\arcsec$, equivalent to $8.3\times6.4$~kpc at 32 Mpc, with a position angle of 323\fdg9. We defined an ellipse enclosing the source with semi-axes $72\arcsec \times 34\arcsec$ centered at $12^{\mathrm h}18^{\mathrm m}13\fs2, 07\arcdeg21\arcmin48\arcsec$ and with the same position angle as the restoring beam, chosen to enclose the maximum flux. Flux measurements on the spectrum extracted from this elliptical region gave a value of $0.20 \pm 0.03$~\jykms.

\begin{figure}
    \plotone{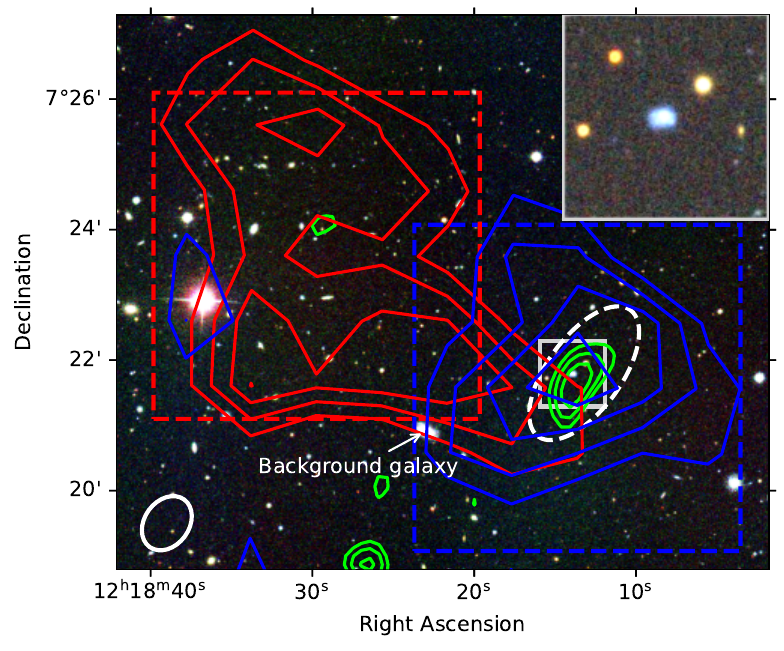}
    \caption{\hi\ contours of AGESVC1~231 from 0.075~\jybmkms\ in steps of 0.025~\jybmkms\ (red = AGES 1945--2080~\kms; green = VLA 1875--1945~\kms; blue = AGES 1875--1945~\kms) overlaid on a color (RGB=$irg$) SDSS image. The VLA beam (white) is shown to the lower left; the AGES beam is circular with a diameter of 3\farcm5, giving it 20 times the area of the VLA beam. The dashed regions show the areas where the `tail' (red) and `body' (blue) AGES spectra and the VLA spectrum were extracted (see figure \ref{fig:VC1_231_spectra}). An optical image from the Legacy Survey centered on the optical counterpart, covering the $1\arcmin\times1\arcmin$ area indicated by the gray box, is inset in the top right.}
    \label{fig:VC1_231_contours}
\end{figure}

The flux detected is consistent with that seen in the original AGES detection, but with a much narrower velocity width of $60\pm20$~\kms\ at the 20 percent level. The VLA observations enabled us to identify the \hi\ source with a small, very blue dwarf galaxy, \object{SDSS J121813.93+072147.3} (SDSS: $g = 18.62\pm0.01$; $g-r = 0.03\pm0.02$), which can be seen in figure \ref{fig:VC1_231_contours} about 10\arcsec\ north of the peak of the VLA \hi\ distribution. The brighter galaxy south east of the VLA detection has an SDSS spectroscopic redshift of $z = 0.064$, well beyond the Virgo cluster.

\begin{figure*}
    \plottwo{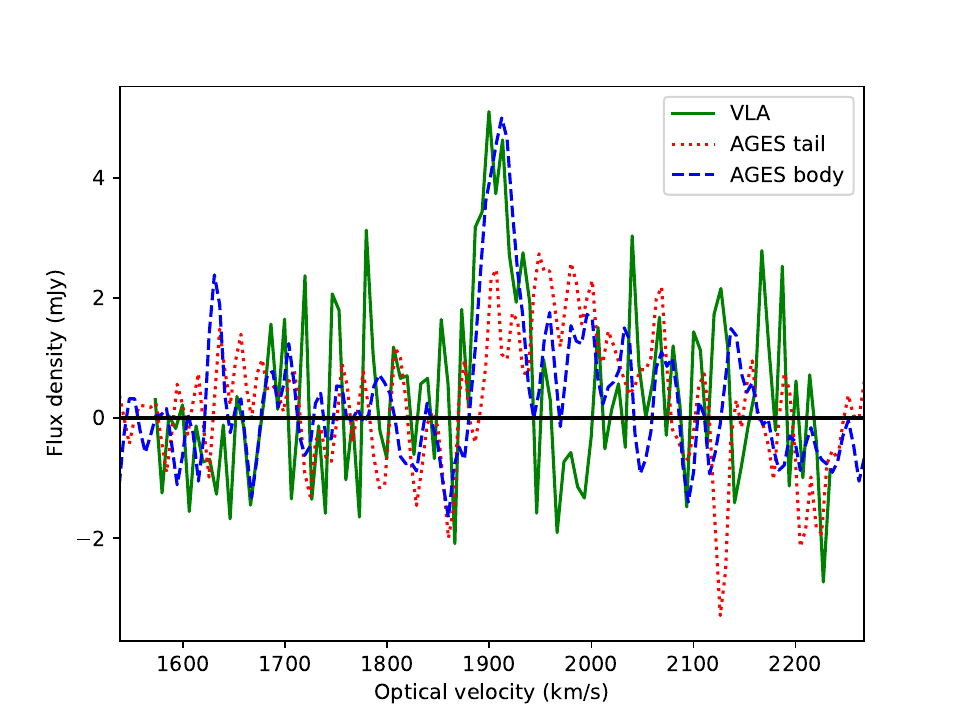}{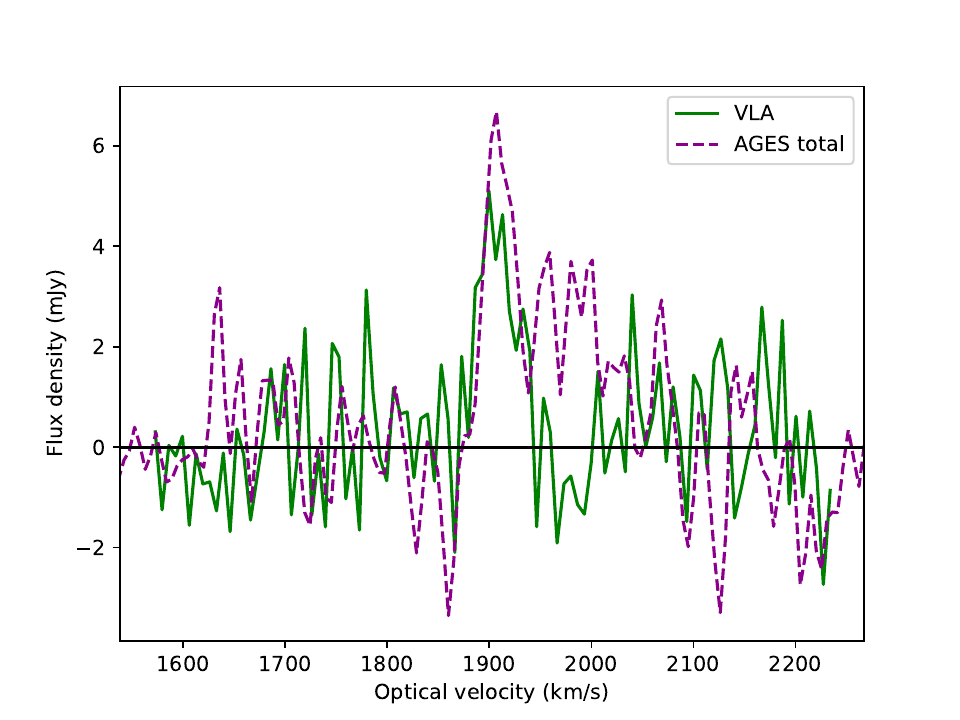}
    \caption{\hi\ spectra of AGESVC1~231 showing left: VLA (green solid line) and AGES body  (blue dashed line) and tail (red dotted line) regions, and right: VLA (green solid line) and AGES total (magenta dashed line).}
    \label{fig:VC1_231_spectra}
\end{figure*}

Following \citet{2019AJ....158..121M}'s treatment of NGC~4522, we created moment maps covering the main peak, with similar velocity limits to the VLA detection, and the velocity range where AGES saw \hi\ but the VLA did not. As can be seen in figure \ref{fig:VC1_231_contours}, an extended tail of lower column-density emission was detected. AGES spectra from this tail region and the region around the galaxy, along with the VLA spectrum of the galaxy, can be seen in figure \ref{fig:VC1_231_spectra}. These were formed using ``yaxis = sum'' in mbspect and then beam-corrected to give the flux density.

Assuming the VLA has detected all of the high column-density gas in the galaxy, which appears to be the case from figure \ref{fig:VC1_231_spectra} (left), the difference between the VLA and AGES spectra will be the low column-density stripped gas. To investigate this, we combine the spectra from the two AGES regions (subtracting the spectrum from the area where the regions overlap to avoid double-counting this). The results are shown in figure \ref{fig:VC1_231_spectra} (right).

This analysis shows that around 60 percent of the \hi\ in this system is in the stripped gas, most of  which was not included in the original AGES measurement and thus leads to a large increase in the measured \hi\ mass. At the 32 Mpc distance assumed for AGESVC1~231, the total detected \hi\ flux from AGES of $0.51\pm0.06$~\jykms\ equates to an \hi\ mass of $12.3\pm1.5 \times 10^7~\nom{M}$, with $4.8\pm0.7 \times 10^7~\nom{M}$ detected by the VLA in the galaxy and thus $7.5\pm1.7 \times 10^7~\nom{M}$ in the stripped gas. From the SDSS optical measurements we derive $L_g = 4.3 \times 10^7~\nom{L}$, and thus $M_{HI}/L_g = 2.9~\nom{M}/\nom{L}$ in the system as a whole and $1.1~\nom{M}/\nom{L}$ for the gas remaining in the galaxy.

\begin{figure*}
\plottwo{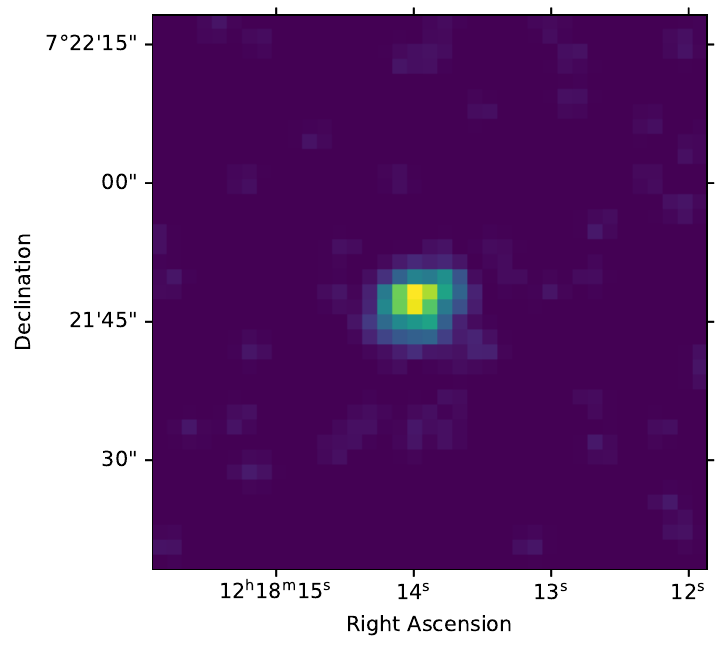}{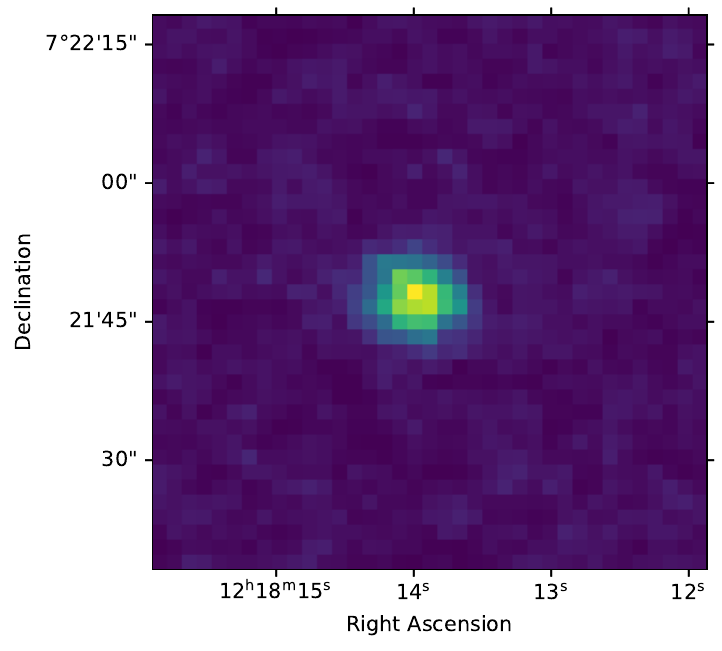}
\plottwo{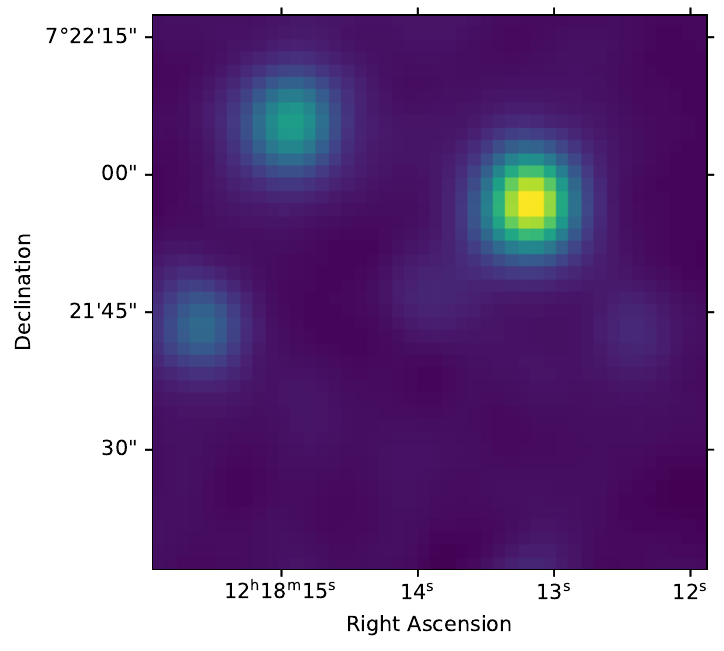}{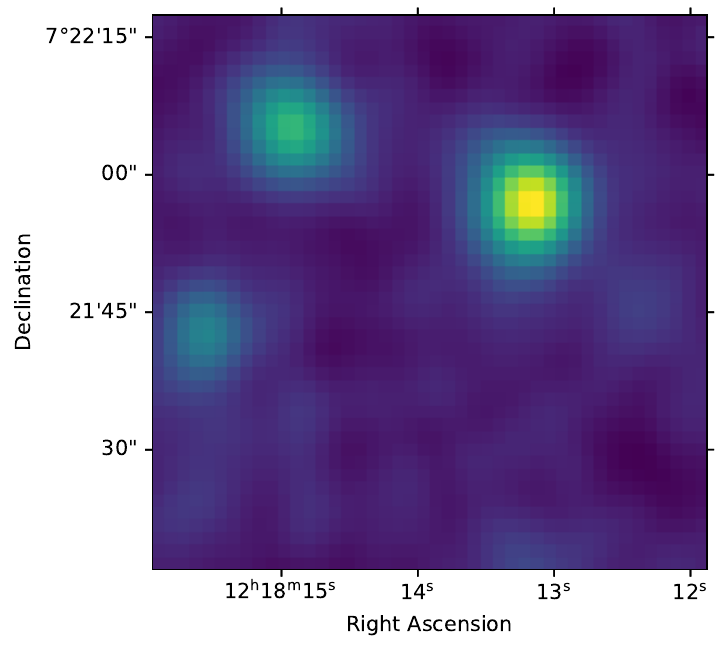}
\caption{Ultraviolet and infrared images centered on the optical coordinates of the counterpart to AGESVC1~231 and covering the same $1\arcmin\times1\arcmin$ area as the gray box and inset image in figure \ref{fig:VC1_231_contours}. Upper left: GALEX FUV band. Upper right: GALEX NUV band. Lower left: AllWISE W1 band. Lower right: AllWISE W2 band.}
\label{VC1_231_UVIR}
\end{figure*}

The counterpart identified here is the same as the optical source associated with AGESVC1~231 by \citet{2012MNRAS.423..787T}, where it was considered an unlikely counterpart (although the best available) due to its very blue color and compact size, in contrast to the high velocity width measured for the \hi\ source. GALEX data shows significant UV emission associated with this source, indicating ongoing star formation. Its compact size and blue color make it similar in appearance to the `blueberry' galaxies identified by \citet{2017ApJ...847...38Y}; however, its extinction-corrected optical color is much less extreme, at $g-r = 0.02\pm0.03$ compared to the $<-0.5$ of their blueberry galaxies, while the 95\% confidence limit on its near infrared color of $W1-W2 < 0.49$, based on the detection in the W1 band and the 95\% confidence upper limit in the W2 band in the AllWISE Catalog (where the fit has a low, ($1.1\sigma$ significance), is consistent with the generally red infrared colors found for star-forming blue dwarf galaxies by \citet{2024A&A...688A.159K} but not with the $W1-W2 > 0.5$ they found for the majority of blueberry galaxies. UV and IR images of the counterpart  are shown in figure \ref{VC1_231_UVIR}

\begin{figure}
    \plotone{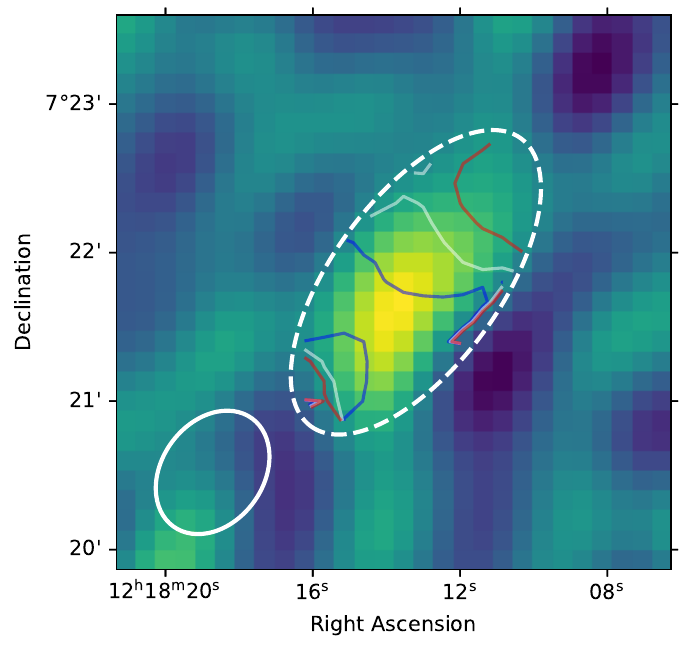}
    \caption{Integrated VLA intensity (moment 0) map of AGESVC1~231 overlaid with contours from the VLA intensity-weighted velocity (moment~1) map at 1890~\kms\ (blue), 1900~\kms\ (white) and 1910~\kms\ (red). The enclosing elipse, which is the same as the dashed elipse in figure \ref{fig:VC1_231_contours}, shows the region over which the \hi\ flux and spectrum were measured and the moment 1 map calculated. The VLA beam is shown to the lower left.}
    \label{fig:A231_mom0_mom1}
\end{figure}

The ellipse used to define the source region is more elongated than the restoring beam, indicating some extension along the major axis. The moment 1 map (contours on figure \ref{fig:A231_mom0_mom1}) shows a slight change in velocity along the major axis but not well-ordered rotation. The peak of the source is at the lowest intensity-weighted velocity with the intensity-weighted velocity getting higher towards both ends of the source. This is consistent with stripping of gas towards higher velocities. The higher velocity width seen in the AGES data is due to gas at higher recessional velocities that is not seen with the VLA, adding further weight to this possibility.

Although the region where AGESVC1~231 is located is quite distant in projection from the center of the Virgo cluster and so might not be expected to have strong ram-pressure effects based on its position relative to the main cluster, the environmental properties of the 32 Mpc sub-cluster are less well known. Evidence for ram pressure in this 32 Mpc sub-cluster has been previously found by \citet{2020AJ....159..218T}.

\subsection{AGESVC1~257}
\begin{figure}
    \plotone{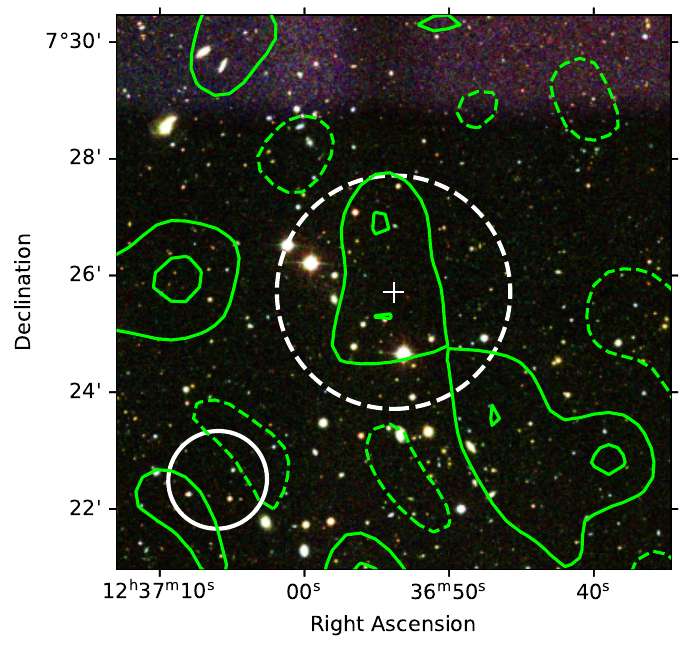}
    \caption{VLA \hi\ contours in the field of AGESVC1~257 at -2, -1, 1, 2$\sigma$ (negative contours shown dashed) overlaid on a color (RGB=$irg$) SDSS image. The region in which the spectrum shown in figure \ref{fig:VC1_257_spectra} was measured is shown by the dashed white circle and the VLA beam is shown by the solid white ellipse to the lower left. The \hi\ position from our remeasurement of the AGES data is shown by the white cross.}
    \label{fig:VC1_257_contours}
\end{figure}

\begin{figure}
  \plotone{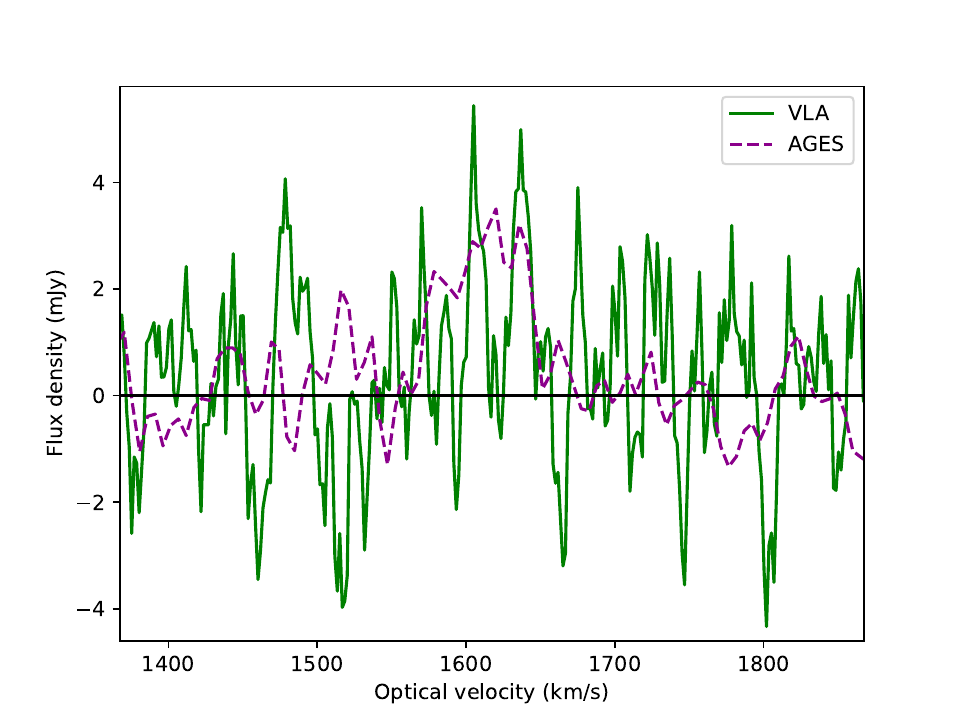}
    \caption{\hi\ spectra at the position of AGESVC1~257 from the VLA (green solid line) and AGES (magenta dashed line). VLA data boxcar smoothed over 6 channels to a velocity resolution of 10~\kms; AGES data presented at native 5~\kms\ resolution.}
    \label{fig:VC1_257_spectra}
\end{figure}

AGESVC1~257 was not detected with the VLA. A UV taper of $1.15 {\rm k}\lambda \times 1.1 {\rm k}\lambda$ was applied in {\sc tclean} to give a beam of $102\arcsec \times 100\arcsec$ and a moment map formed in CARTA \citep{2021zndo...3377984C} over the $\Delta V_{50}$ range of 1574--1645~\kms\ from our remeasurement of the AGES cube (see figure \ref{fig:VC1_257_contours}. A spectrum formed with the same 2\arcmin\ radius used for the weaker VLA detections, centered on the remeasured position of the AGES detection, is shown in figure \ref{fig:VC1_257_spectra}. The measured noise in the moment map gives a $3\sigma$ column-density limit for the non-detection of $1.1 \times 10^{19} {\rm\ cm}^{-2}$, compared to a $3\sigma$ level of $1.2 \times 10^{18} {\rm\ cm}^{-2}$ in the AGES cube. It is therefore likely that a sufficient fraction of the gas detected in AGES is below the VLA column-density limit to render the source undetectable here. The $3\sigma$ limit on the \hi\ mass in the VLA observations, assuming a source size the same as the AGES beam, is more than twice the \hi\ mass detected with AGES. From the VLA column-density limit and the AGES \hi\ flux density, the minimum size (FWHM) of a circular Gaussian \hi\ distribution that matched these limits would be 143\arcsec.

\newpage
\subsection{AGESVC1~258}

\begin{figure}
    \plotone{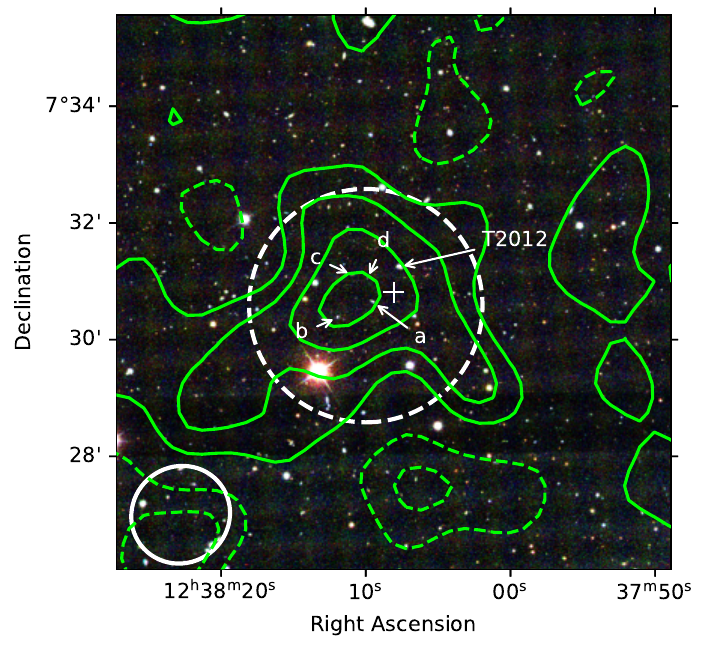}
    \caption{VLA \hi\ contours of AGESVC1~258 at -2, -1, 1, 2, 3, 4 $\sigma$ (negative contours shown dashed) overlaid on a color (RGB=$irg$) SDSS image. The region in which the \hi\ flux was measured is shown by the dashed white circle and the uv-tapered VLA beam is shown by the solid white ellipse to the lower left. The \hi\ position from our remeasurement of the AGES data is shown by the white cross. SDSS sources within the $4\sigma$ contour are labeled `a' to `d' from brightest to faintest, and the candidate identified by \citet{2012MNRAS.423..787T} is labeled `T2012'.}
    \label{fig:VC1_258_contours}
\end{figure}

\begin{figure}
    \plotone{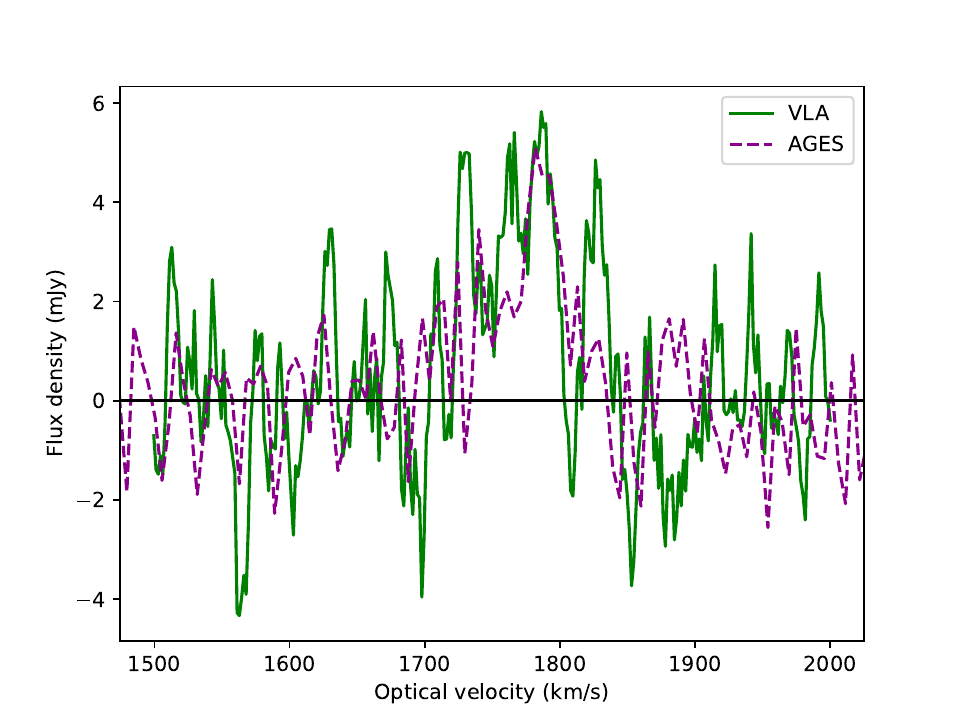}
    \caption{\hi\ spectra of AGESVC1~258 from the VLA (green solid line) and AGES (magenta dashed line). VLA data boxcar smoothed over 6 channels to a velocity resolution of 10~\kms; AGES data presented at native 5~\kms\ resolution.}
    \label{fig:VC1_258_spectra}
\end{figure}

AGESVC1~258 was marginally detected in the VLA data. Imaging with a UV taper of $1.2 {\rm k}\lambda \times 1.2 {\rm k}\lambda$ gave a $103\arcsec\times99\arcsec$ beam and a noise per channel of 2.3 mJy/beam, resulting in a 4$\sigma$ (peak) detection in a moment map formed over a fairly wide range of 1725--1805~\kms, with the peak at $12^{\mathrm h}38^{\mathrm m}11^{\mathrm s}, 07\arcdeg30\arcmin40\arcsec$.  The moment map is shown overlain as contours on an SDSS image in figure \ref{fig:VC1_258_contours}. The flux density measured on the spectrum extracted from a 2\arcmin\ radius aperture (centered at $12^{\mathrm h}38^{\mathrm m}10^{\mathrm s}, 07\arcdeg30\arcmin35\arcsec$) enclosing the source is $0.29\pm0.05$~\jykms, consistent with the flux measured in the AGES cube. The detection seen with the VLA is also coincident with the AGES detection in velocity, giving confidence that it is real, as shown in figure \ref{fig:VC1_258_spectra} with the VLA data smoothed to 10~\kms\ resolution to reach a noise level of 1.5 mJy. Splitting the velocity range into two parts fails to reveal any ordered structure to the gas in AGESVC1~258, although the signal to noise ratio is obviously low.

There are four objects classified as galaxies by the SDSS (labeled `a' to `d' on figure \ref{fig:VC1_258_contours} in order of decreasing $g$-band brightness) within the $4\sigma$ contour, also visible on the DESI Legacy Survey, all near this contour rather than in the center. None have SDSS spectroscopy. The brightest, at $g=20.76$ and $g-r=0.53$, appears very elongated and so seems likely to be a background inclined spiral. The second brightest, at $g=20.97$, $g-r=0.27$, would give an $M_{HI}/L_g$ of $13.6~\nom{M}/\nom{L}$. The third brightest, at $g=21.87$, $g-r=1.04$, is much redder and has less than half the luminosity, making it likely to be a background galaxy. The fourth is fainter still, at $g=23.05$, $g-r=0.54$ which would give an $M_{HI}/L_g$ of $92~\nom{M}/\nom{L}$. While the second and fourth candidates would have much higher gas-to-light ratios than the ``(almost) dark'' ALFALFA galaxies of \citet{2017ApJ...842..133L}, AGESVC1~274 (see below) and all of the sources with identified optical counterparts in \citet{2012MNRAS.423..787T}, they would still have lower gas fractions than those found for BC6 (with a gas-to-stellar-mass ratio of around 3,000), the spectroscopically-confirmed optical counterpart to the ALFALFA Virgo 7 cloud complex \citep{2024ApJ...966L..15J} and the gas-to-stellar-mass ratio of around 1,100 found for BC17, the proposed optical counterpart of AGESVC1~266 \citep{2025ApJ...983....2D}. Thus these optical sources cannot be entirely ruled out as optical counterparts; however, given the lack of spectroscopic data or other supporting evidence, we do not identify either of these as the optical counterpart of AGESVC1~258. The candidate identified by \citet{2012MNRAS.423..787T} lies around 55\arcsec\ northwest of the \hi\ peak, within the $3\sigma$ contour, and is similar in appearance on the Legacy Survey images to the counterpart to AGESVC1~231, although about half a magnitude fainter. This is a similar offset to the counterpart identified by \citet{2025ApJ...983....2D} to AGESVC1~266 and so cannot be ruled out, although the optical image does not resemble a ram-pressure dwarf making it an unlikely counterpart for this \hi\ detection.

\subsection{AGESVC1~262}
\begin{figure}
    \plotone{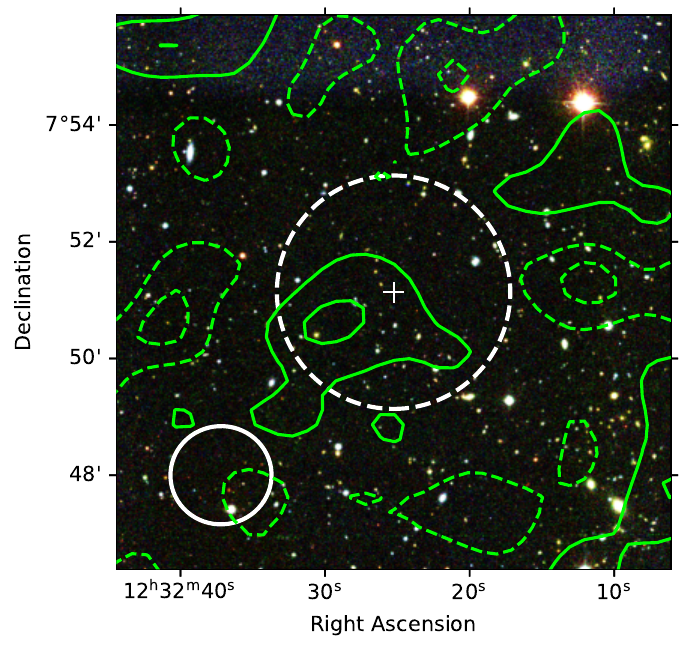}
    \caption{VLA \hi\ contours in the field of AGESVC1~262 at -2, -1, 1, 2$\sigma$ (negative contours shown dashed) overlaid on a color (RGB=$irg$) SDSS image. The region in which the spectrum shown in figure \ref{fig:VC1_262_spectra} was measured is shown by the dashed white circle and the VLA beam is shown by the solid white ellipse to the lower left. The \hi\ position from our remeasurement of the AGES data is shown by the white cross.}
    \label{fig:VC1_262_contours}
\end{figure}

\begin{figure}
  \plotone{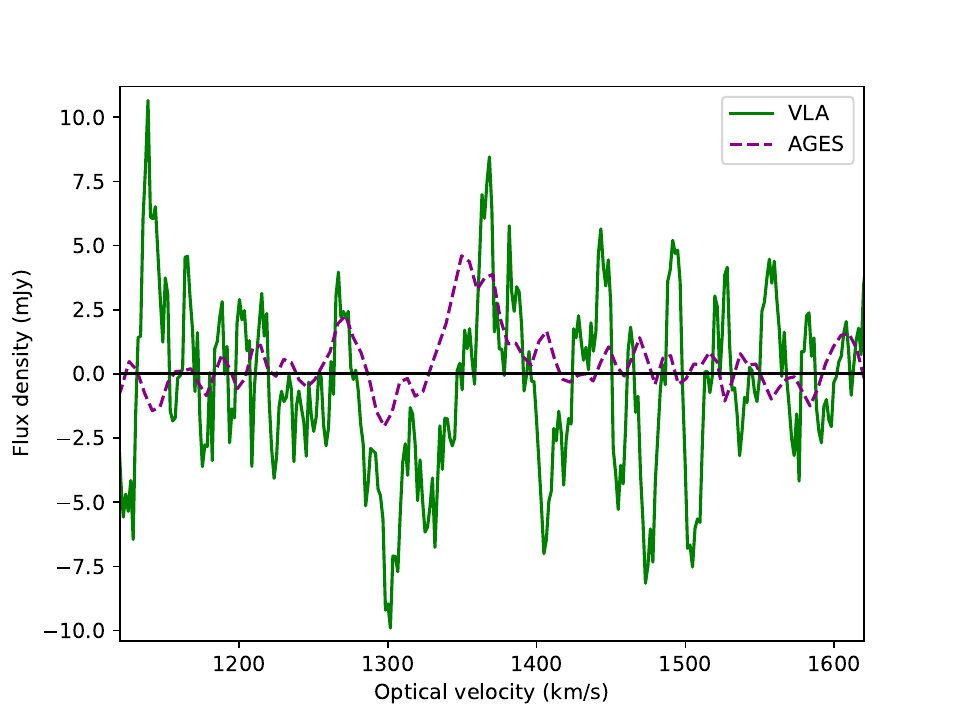}
    \caption{\hi\ spectra at the position of AGESVC1~262 from the VLA (green solid line) and AGES (magenta dashed line). VLA data boxcar smoothed over 6 channels to a velocity resolution of 10~\kms; AGES data presented at native 5~\kms\ resolution.}
    \label{fig:VC1_262_spectra}
\end{figure}

AGESVC1~262 was not detected with the VLA. A UV taper of $1.2 {\rm k}\lambda \times 1.1 {\rm k}\lambda$ was applied in {\sc tclean} to give a beam of $104\arcsec \times 103\arcsec$ and a moment map formed in CARTA \citep{2021zndo...3377984C} over the $\Delta V_{50}$ range of 1341--1375~\kms\ from our remeasurement of the AGES cube (see figure \ref{fig:VC1_262_contours}. A spectrum formed with the same 2\arcmin\ radius used for the weaker VLA detections, centered on the remeasured position of the AGES detection, is shown in figure \ref{fig:VC1_262_spectra}. The measured noise in the moment map gives a $3\sigma$ column-density limit for the non-detection of $8.9 \times 10^{18} {\rm\ cm}^{-2}$, compared to a $3\sigma$ level of $8.4 \times 10^{17} {\rm\ cm}^{-2}$ in the AGES cube. As with AGESVC1-257, it is therefore likely that a sufficient fraction of the gas detected in AGES is below the VLA column-density limit to render the source undetectable here. As in that source, the $3\sigma$ limit on the \hi\ mass in the VLA observations here, assuming a source size the same as the AGES beam, is more than twice the \hi\ mass detected with AGES. From the VLA column-density limit and the AGES \hi\ flux density, the minimum size (FWHM) of a circular Gaussian \hi\ distribution that matched these limits would be 142\arcsec.

\subsection{AGESVC1~266}

\begin{figure}
    \plotone{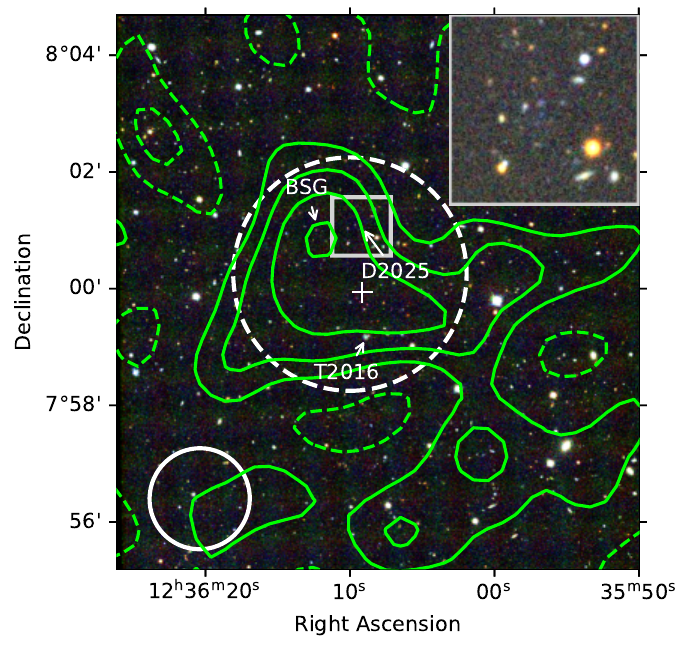}
    \caption{VLA \hi\ contours of AGESVC1~266 at -2, -1, 1, 2, 3, 4 $\sigma$ (negative contours shown dashed) overlaid on a color (RGB=$irg$) SDSS image. The region in which the \hi\ flux was measured is shown by the dashed white circle and the VLA beam is shown by the solid white ellipse to the lower left.  An optical image from the Legacy Survey centered on the optical counterpart identified by \citet{2025ApJ...983....2D}, covering the area shown by the gray box, is inset in the top right. The \hi\ position from our remeasurement of the AGES data is shown by the white cross. The brightest SDSS galaxy within the $4\sigma$ contour is labeled `BSG', the candidate identified by \citet{2016MNRAS.461.3001T} is labeled `T2016', and the counterpart identified by \citet{2025ApJ...983....2D} is labeled `D2025'.}
    \label{fig:VC1_266_contours}
\end{figure}

AGESVC1~266 was marginally detected in the VLA data. Imaging with a UV taper of $1.35 {\rm k}\lambda \times 1.05 {\rm k}\lambda$ gave a $104\arcsec\times103\arcsec$ beam and a noise per channel of 2.4 mJy/beam, resulting in a 4$\sigma$ (peak) detection in a moment map formed over a very narrow range of 1684.5--1694.5~\kms, with the peak at $12^{\mathrm h}36^{\mathrm m}12^{\mathrm s}, 08\arcdeg00\arcmin50\arcsec$. The flux measured on the spectrum extracted from a 2\arcmin\ radius aperture (centered at $12^{\mathrm h}36^{\mathrm m}10^{\mathrm s}, 08\arcdeg00\arcmin15\arcsec$) enclosing the source is $0.11\pm0.04$~\jykms. No optical counterpart was identified for this source in this work. The moment map is shown overlain as contours on an SDSS image in figure \ref{fig:VC1_266_contours}. There are a number of optical sources within the $4\sigma$ contour identified as galaxies by the SDSS or found in examination of the DESI Legacy Survey. None of these are obvious counterparts, with the brightest (and furthest from the center) having $g=22.87$. These would all have extreme gas-to-light ratios, so while we cannot entirely rule any of these objects out we do not identify any of them as the optical counterpart to AGESVC1~266.

While this study was being prepared for publication, \citet{2025ApJ...983....2D} identified a `blue blob', BC17, at $12^{\mathrm h}36^{\mathrm m}09\fs2, +08\arcdeg01\arcmin04\arcsec$, as the optical counterpart of AGESVC1~266 on the basis of H$\alpha$ spectroscopy. This object lies on the $3\sigma$ contour of our map, around 45\arcsec\ northeast of the \hi\ peak. They identify this as having a stellar mass of $1.1\times10^4$~\nom{M} and thus a gas-to-stellar-mass ratio of around 1,100. The candidate identified by \citet{2016MNRAS.461.3001T} lies further south at a greater distance from the \hi\ peak (although closer to the AGES position for the cloud); no redshift is available for this source but \citet{2025ApJ...983....2D} suggest it is likely to be a background galaxy on the basis of its appearance in high-quality optical imaging. The identified counterpart is possibly a ram-pressure dwarf, similar to those identified by \citet{2022ApJ...935...51J}. The separation of the optical counterpart from the \hi\ peak, which is where star formation would appear most likely, potentially points to on-going ram-pressure effects that have displaced the gas from the optical counterpart. 

\begin{figure}
    \plotone{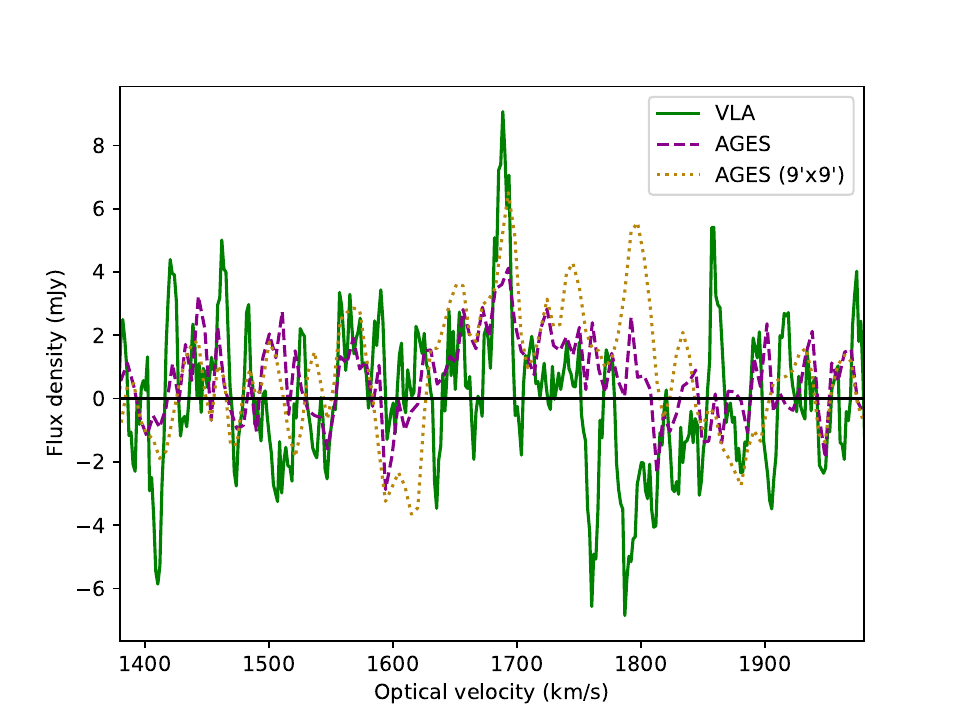}
    \caption{\hi\ spectra of AGESVC1~266 from the VLA (green solid line), AGES (magenta dashed line), and AGES data over a larger $9\arcmin\times9\arcmin$ region (gold dotted line). VLA data boxcar smoothed over 6 channels to a velocity resolution of 10~\kms; AGES data presented at native 5~\kms\ resolution.}
    \label{fig:VC1_266_spectra}
\end{figure}

The \hi\ flux detected with the VLA represents only 35 percent of the AGES flux, and is over a much narrower velocity range, indicating that the majority of the \hi\ in this object is in low column-density gas that could not be detected with the VLA. This would be consistent with it being a dispersing cloud removed from its host by tidal interactions or ram-pressure stripping. The detection seen with the VLA is also coincident with the peak of the AGES data in velocity, giving confidence that it is real, as shown in figure \ref{fig:VC1_266_spectra} with the VLA data smoothed to 10~\kms\ resolution to reach a noise level of 2.1 mJy. However, the VLA spectrum shows a peak flux density about twice as high as that seen with AGES, although this is only about $2\sigma$ higher as the  peak is only a little over $4\sigma$, raising the possibility that this cloud is only visible in the VLA data due to noise enhancing a real signal. Using a $9\arcmin\times 9\arcmin$ region for the measurement of the AGES flux rather than the standard $5\arcmin\times 5\arcmin$ box makes up most of the missing flux in the peak (gold dotted line in figure \ref{fig:VC1_266_spectra}), but the total flux measurement is then affected by confusion with the nearby VCC 1675 (AGESVC1~265) at 1799~\kms.

VCC 1675 lies 6\farcm2 to the northeast of the peak of AGESVC1~266, with a separation of around 100~\kms\ in velocity. This makes it a possible origin for the AGESVC1~266 cloud, either as a product of ram-pressure stripping or harassment. Further, VCC 1675 was identified by \citet{2012MNRAS.423..787T} as very gas deficient, with a deficiency of 1.55, and so has almost certainly undergone gas removal in the past; the gas in AGESVC1~266 would only account for a small fraction (around 2\%) of the gas thought to have been removed from VCC 1675, implying that if this cloud is the remnant of that gas then most has already dispersed into the intra-cluster medium.

\subsection{AGESVC1~274}

AGESVC1~274 is thought to be in the main body of the Virgo cluster at 17 Mpc \citep{2012MNRAS.423..787T}. The original AGES measurements found an \hi\ flux of $0.107 \pm 0.30$~\jykms\ and a velocity width at 20\% of the peak of $35\pm6$~\kms.

\begin{figure}
    \plotone{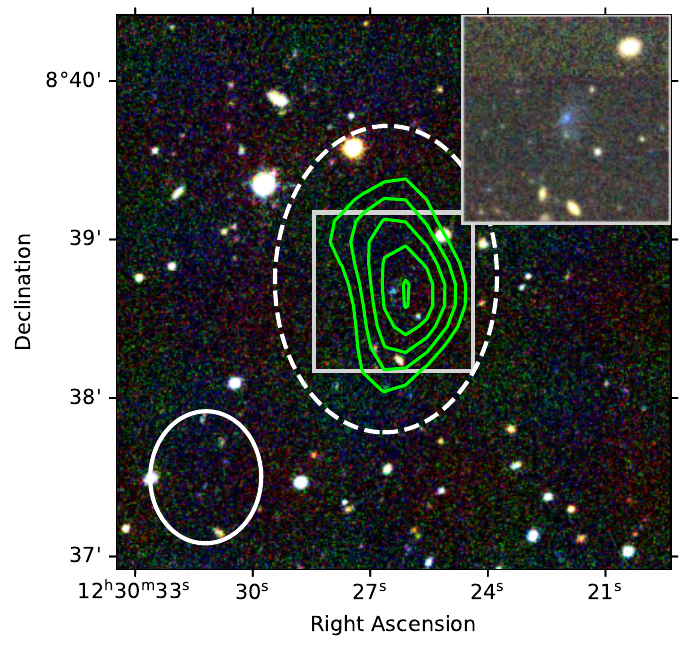}
    \caption{VLA \hi\ contours of AGESVC1~274 from 0.075~\jybmkms\ in steps of 0.025~\jybmkms\ (green) overlaid on a color (RGB=$irg$) SDSS image. The VLA beam is shown by the white ellipse to the lower left. The dashed white ellipse shows the region over which the spectrum in Figure \ref{fig:VC1_274_spectra} was extracted.  An optical image from the Legacy Survey centered on the optical counterpart, covering the area shown by the gray box, is inset in the top right.}
    \label{fig:VC1_274_contours}
\end{figure}

This source was detected with the VLA and a clean map made at the original 1.7~\kms\ velocity resolution of the data, from which a moment 0 map was formed over 1275 to 1315~\kms, which can be seen in figure \ref{fig:VC1_274_contours}. The restoring beam was $50\arcsec\times42\arcsec$, equivalent to $4.1\times3.4$~kpc at 17 Mpc, with a position angle of 358\fdg7. We defined an ellipse enclosing the source with semi-axes $58\arcsec \times 42\arcsec$ centered at $12^{\mathrm h}30^{\mathrm m}26\fs6, 08\arcdeg38\arcmin45\arcsec$ and with the same position angle as the restoring beam, chosen, as for AGESVC1~231, to enclose the maximum \hi\ flux. Flux measurements on the spectrum extracted from this elliptical region gave a value of $0.21\pm0.03$~\jykms. 

\begin{figure}
    \plotone{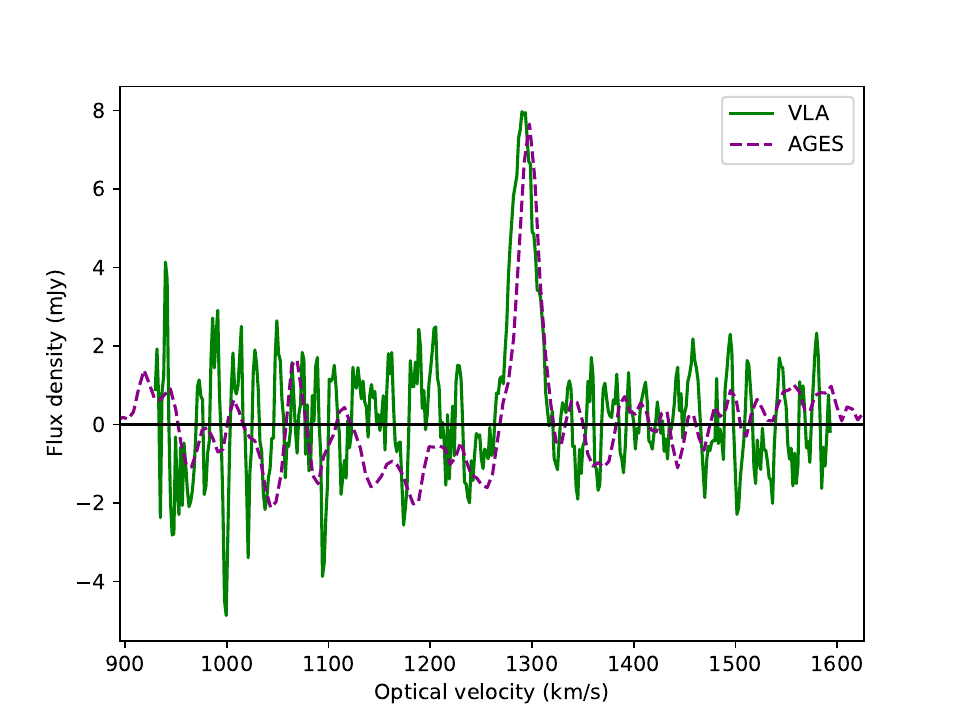}
    \caption{\hi\ spectra of AGESVC1~274 from the VLA (green solid line) and AGES (magenta dashed line), showing good correspondence between the two.}
    \label{fig:VC1_274_spectra}
\end{figure}

The flux detected is higher than seen in the original AGES measurements but consistent with our remeasurement of the AGES data (using the improved VLA position as a starting point) that found a flux of $0.18\pm 0.02$~\jykms. Very similar velocity widths were found at 20\% of the peak in both cases: $35\pm6$~\kms\ in the remeasured AGES data and $39\pm9$~\kms\ in the VLA data. The remeasured AGES spectrum and the VLA spectrum are shown in figure \ref{fig:VC1_274_spectra}; it can be seen that there is good agreement. It is possibly significant that this source also does not have the large difference between $\Delta\mathrm{V}_{50}$ and $\Delta\mathrm{V}_{20}$ in the AGES data that is seen in the other detections, indicating a relatively undisturbed \hi\ profile.

\begin{figure*}
\plottwo{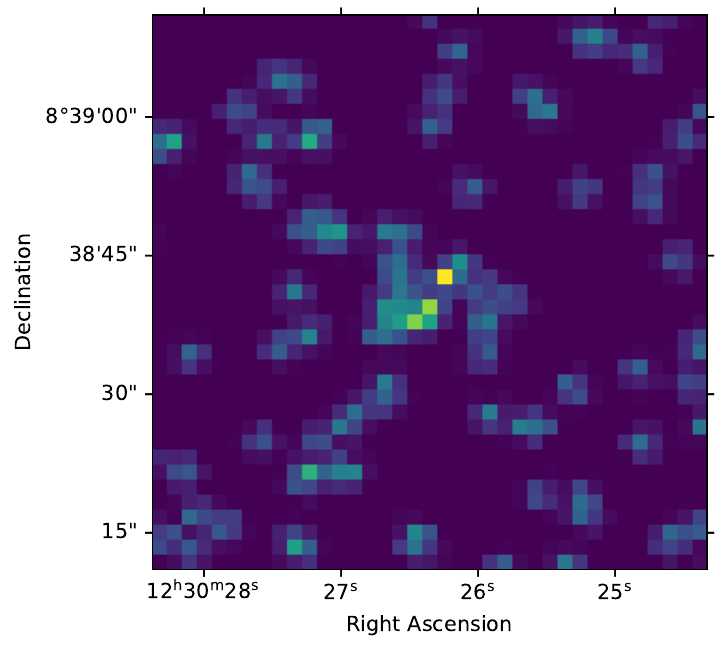}{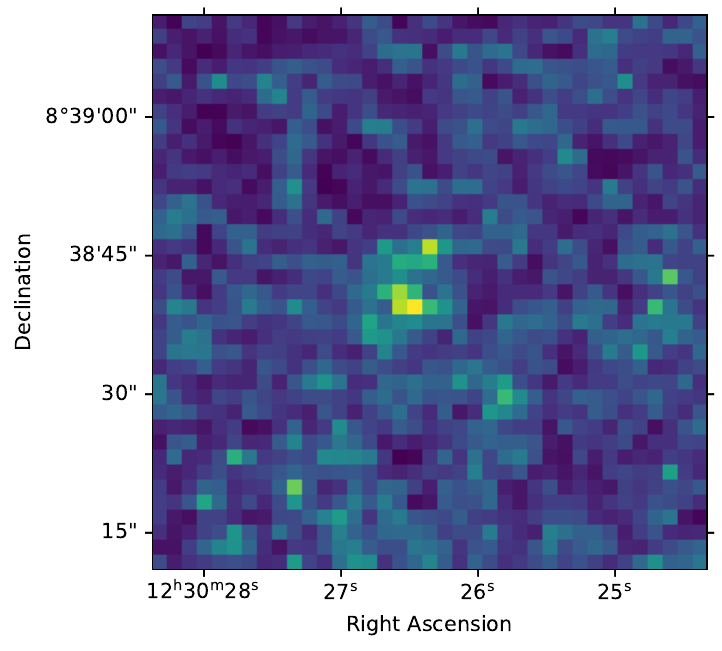}
\plottwo{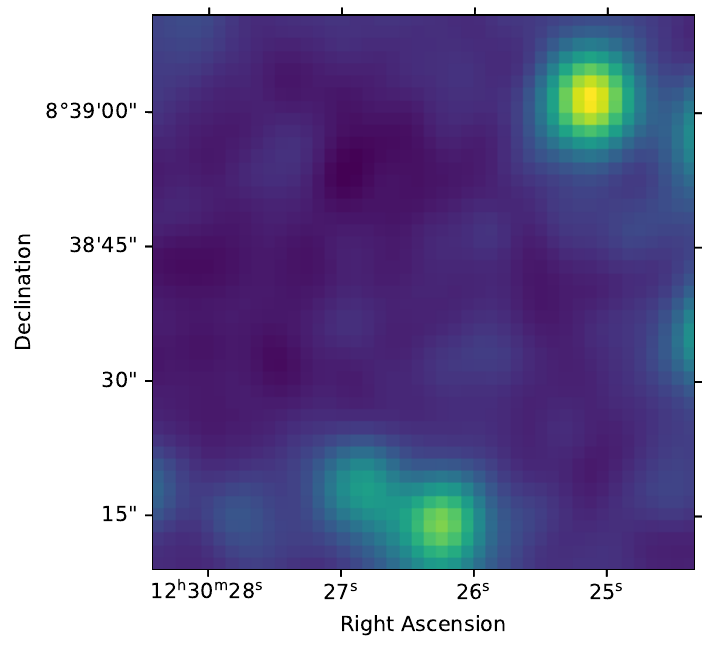}{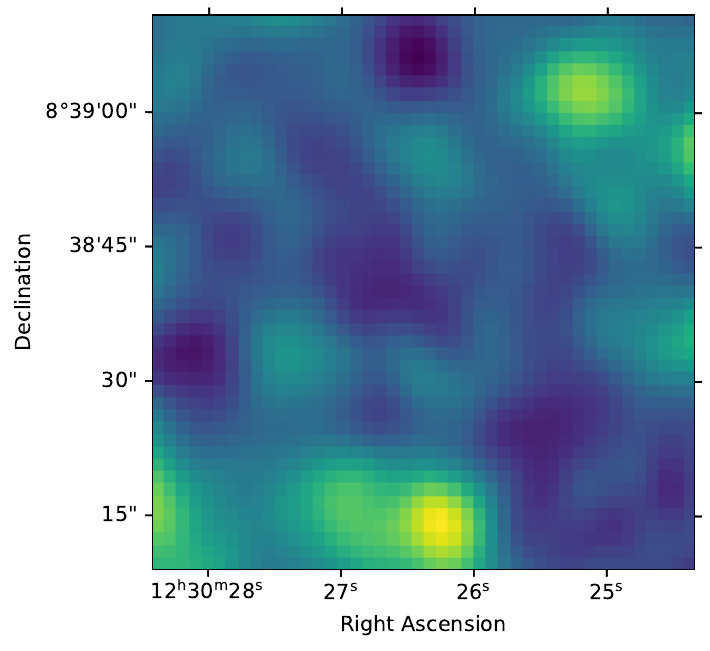}
\caption{Ultraviolet and infrared images centered on the optical coordinates of the counterpart to AGESVC1~274. Upper left: GALEX FUV band. Upper right: GALEX NUV band. Lower left: AllWISE W1 band. Lower right: AllWISE W2 band.}
\label{VC1_274_UVIR}
\end{figure*}

A faint blue optical counterpart, \object{SDSS J123026.40+083840.2}, was identified about 5\arcsec\ east of the peak of the \hi\ moment map, shown in figure \ref{fig:VC1_274_contours}. This is the same source identified as a possible counterpart by \citet{2016MNRAS.461.3001T}, which is being targeted as part of an optical follow-up program. In many ways, this source appears to be a fairly standard dwarf galaxy composed of high column-density \hi\ with no obvious signs of disturbance. As with the counterpart to AGESVC1~231, this source is detected by GALEX, indicating on-going star formation. However, the optical counterpart is low surface-brightness and very faint (SDSS: $g=20.73\pm0.11$, $g-r = 0.42\pm0.16$),  giving it $L_g=1.7\pm0.2\times10^6~\nom{L}$ and thus a very high $M_{HI}/L_g$ of $8.3\pm1.4~\nom{M}/\nom{L}$. This is substantially higher than the ``(almost) dark'' ALFALFA galaxies of \citet{2017ApJ...842..133L} and higher than all but VCC~203/AGESVC1~252 from the sources with identified optical counterparts in \citet{2012MNRAS.423..787T}. The counterpart is also detected in both the NUV and FUV ultraviolet bands by GALEX but is not detected in the infrared by AllWISE in either the W1 band or the W2 band. The UV and IR images are shown in figure \ref{VC1_274_UVIR}.

While this work was being prepared for publication, the same counterpart was independently confirmed by \citet{2025ApJ...983....2D} and identified as a candidate `blue blob'. However, contrary to \citet{2025ApJ...983....2D}'s conclusion that ram pressure is the most likely formation mechanism for such objects, the \hi\ data here show that all the gas is in a compact object, with no significant low column-density gas seen in AGES but not detected by the VLA. This points to AGESVC1~274 being a primordial structure rather than a condensation of ram-pressure stripped gas. This may indicate that some optical sources identified as blue blobs are formed by ram-pressure stripping and others are primordial objects -- high resolution \hi\ imaging of the high column-density gas, ideally combined with single-dish measurements to detect surrounding low column-density gas, may be critical in distinguishing between these scenarios.

\section{Discussion and conclusions} \label{sec:discussion}

Our observations have uncovered a great diversity in the population of nearly-dark clouds from \citet{2012MNRAS.423..787T}:
\begin{enumerate}
\item AGESVC1~274 has all of its gas in a compact object that can be identified with an optical counterpart.
\item AGESVC1~231 has around 40 percent of its gas in a compact object that can be identified with an optical counterpart and the remainder in a tail that is not detected with the VLA.
\item AGESVC1~258 has all of its gas in an extended, low column-density object with a peak column-density of $1.6\times10^{19}{\rm\ cm}^{-2}$ (after smoothing) and a VLA flux consistent with the AGES flux, and a very faint optical counterpart identified separately from this work by \citet{2025ApJ...983....2D}.
\item AGESVC1~266 has all of its gas in an extended, low column-density object with a peak column-density of $4.3\times10^{18}{\rm\ cm}^{-2}$ (after smoothing) and only a third of the gas detected with the VLA, confined to a  narrower velocity range than the gas detected in AGES.
\item AGESVC1~257 and AGESVC1~262 have none of their gas detected with the VLA and upper limits to their column-density (after spatial smoothing) of $5.2\times10^{18}{\rm\ cm}^{-2}$ and $8.9\times10^{18}{\rm\ cm}^{-2}$ respectively, and thus must also be extended objects. The detection of AGESVC1~266 is consistent with these non-detections, as its peak column-density is less than their $3\sigma$ upper limits. 
\end{enumerate}

The first two of these –- AGESVC1~274 and AGESVC1~231 -- appear to be stable, presumably long-lived objects, albeit undergoing ram-pressure stripping in the case of AGESVC1~231. The optical counterparts to these stable objects are faint, gas-rich galaxies that are unusual but do not call for exotic explanations. We do not see any evidence here of long-lived compact pressure-supported turbulent spheres, as suggested by \citet{2016ApJ...824L...7B}. Based on equation 3 in that paper and our remeasured AGES parameters, these would be expected to be Gaussian spheres with sizes similar to or smaller than the untapered VLA beam, making them easily detectable but without any optical counterpart. This is inconsistent with our observations. We also do not see any evidence of stable dark galaxies -- all of our compact \hi\ detections are associated with optical counterparts.

The other four are extended, low column-density objects that seem unlikely to be stable. The one identified optical counterpart \citep[by][]{2025ApJ...983....2D} appears to be a ram-pressure dwarf. However, this is offset from the \hi\ peak of AGESVC1~266 and seems more likely to be the last flicker of star formation from an \hi\ condensation within a dispersing cloud than a stable, self-gravitating system in formation. 

The diversity of objects identified through our VLA observations and the narrower linewidths in our re-analysis of the AGES data point to an increased likelihood that these gas clouds can be explained by conventional ram-pressure stripping and harassment. While all six of these objects looked similar in the original AGES data, they are shown here to not be a homogeneous population, reducing the force of the statistical arguments in \citet{2017MNRAS.467.3648T} that apply to formation by a single mechanism. The narrower velocity widths seen in the VLA data, compared to the AGES data, also increase the frequency with which \hi\ clouds similar to these would be seen in the simulations of \citet{2017MNRAS.467.3648T}.

We therefore conclude that gas removal from galaxies is the most likely origin of the dark \hi\ clouds in the VC1 region that cannot be associated with faint but stable galaxies. Further, it is likely that these objects are relatively short lived. Either we are looking at a region of Virgo that happens to have recently had an infall event that introduced a number of high mass-to-light ratio galaxies and led to the formation of a number of gas clouds, or these clouds are being renewed more frequently than has been thought, although the diversity of clouds here allows for less frequent renewal than was necessary in scenarios where all of the clouds formed a single population. If other regions of Virgo were to show a similar number of dark \hi\ clouds at a similar distance from the cluster core, that would indicate the latter scenario, but if this region turns out to be an outlier that would imply that we have, indeed, caught it at a fortunate time.

%% Also note that the akcnowlodgment environment does not support long amounts of text. If you have a lot of people and institutions to acknowledge, do not use this command. Instead, create a new \section{Acknowledgments}.
%\begin{acknowledgments}
\section*{Acknowledgments}
We thank the anonymous referee for helpful comments that have improved the paper.

The National Radio Astronomy Observatory is a facility of the U.S. National Science Foundation operated under cooperative agreement by Associated Universities, Inc.

This work was supported by the Czech Ministry of Education, Youth and Sports from the Large Infrastructures for Research, Experimental Development and Innovations project LM 2015067; the Czech Science Foundation grant CSF 19-18647S; and the institutional project RVO 67985815.

This work was supported by the Charles University, project GA UK No. 376425.

BD acknowledges funding from the HTM (grant TK202) and the EU Horizon Europe (EXCOSM, grant No. 101159513).

This work has made use of the SDSS. Funding for the SDSS and SDSS-II has been provided by the Alfred P. Sloan Foundation, the Participating Institutions, the National Science Foundation, the U.S. Department of Energy, the National Aeronautics and Space Administration, the Japanese Monbukagakusho, the Max Planck Society, and the Higher Education Funding Council for England. The SDSS website is http://www.sdss.org/.

The Legacy Surveys consist of three individual and complementary projects: the Dark Energy Camera Legacy Survey (DECaLS; Proposal ID \#2014B-0404; PIs: David Schlegel and Arjun Dey), the Beijing-Arizona Sky Survey (BASS; NOAO Prop. ID \#2015A-0801; PIs: Zhou Xu and Xiaohui Fan), and the Mayall z-band Legacy Survey (MzLS; Prop. ID \#2016A-0453; PI: Arjun Dey). DECaLS, BASS and MzLS together include data obtained, respectively, at the Blanco telescope, Cerro Tololo Inter-American Observatory, NSF’s NOIRLab; the Bok telescope, Steward Observatory, University of Arizona; and the Mayall telescope, Kitt Peak National Observatory, NOIRLab. Pipeline processing and analyses of the data were supported by NOIRLab and the Lawrence Berkeley National Laboratory (LBNL). The Legacy Surveys project is honored to be permitted to conduct astronomical research on Iolkam Du’ag (Kitt Peak), a mountain with particular significance to the Tohono O’odham Nation.

NOIRLab is operated by the Association of Universities for Research in Astronomy (AURA) under a cooperative agreement with the National Science Foundation. LBNL is managed by the Regents of the University of California under contract to the U.S. Department of Energy.

This project used data obtained with the Dark Energy Camera (DECam), which was constructed by the Dark Energy Survey (DES) collaboration. Funding for the DES Projects has been provided by the U.S. Department of Energy, the U.S. National Science Foundation, the Ministry of Science and Education of Spain, the Science and Technology Facilities Council of the United Kingdom, the Higher Education Funding Council for England, the National Center for Supercomputing Applications at the University of Illinois at Urbana-Champaign, the Kavli Institute of Cosmological Physics at the University of Chicago, Center for Cosmology and Astro-Particle Physics at the Ohio State University, the Mitchell Institute for Fundamental Physics and Astronomy at Texas A\&M University, Financiadora de Estudos e Projetos, Fundacao Carlos Chagas Filho de Amparo, Financiadora de Estudos e Projetos, Fundacao Carlos Chagas Filho de Amparo a Pesquisa do Estado do Rio de Janeiro, Conselho Nacional de Desenvolvimento Cientifico e Tecnologico and the Ministerio da Ciencia, Tecnologia e Inovacao, the Deutsche Forschungsgemeinschaft and the Collaborating Institutions in the Dark Energy Survey. The Collaborating Institutions are Argonne National Laboratory, the University of California at Santa Cruz, the University of Cambridge, Centro de Investigaciones Energeticas, Medioambientales y Tecnologicas-Madrid, the University of Chicago, University College London, the DES-Brazil Consortium, the University of Edinburgh, the Eidgenossische Technische Hochschule (ETH) Zurich, Fermi National Accelerator Laboratory, the University of Illinois at Urbana-Champaign, the Institut de Ciencies de l’Espai (IEEC/CSIC), the Institut de Fisica d’Altes Energies, Lawrence Berkeley National Laboratory, the Ludwig Maximilians Universitat Munchen and the associated Excellence Cluster Universe, the University of Michigan, NSF’s NOIRLab, the University of Nottingham, the Ohio State University, the University of Pennsylvania, the University of Portsmouth, SLAC National Accelerator Laboratory, Stanford University, the University of Sussex, and Texas A\&M University.

BASS is a key project of the Telescope Access Program (TAP), which has been funded by the National Astronomical Observatories of China, the Chinese Academy of Sciences (the Strategic Priority Research Program “The Emergence of Cosmological Structures” Grant \# XDB09000000), and the Special Fund for Astronomy from the Ministry of Finance. The BASS is also supported by the External Cooperation Program of Chinese Academy of Sciences (Grant \# 114A11KYSB20160057), and Chinese National Natural Science Foundation (Grant \# 12120101003, \# 11433005).

The Legacy Survey team makes use of data products from the Near-Earth Object Wide-field Infrared Survey Explorer (NEOWISE), which is a project of the Jet Propulsion Laboratory/California Institute of Technology. NEOWISE is funded by the National Aeronautics and Space Administration.

The Legacy Surveys imaging of the DESI footprint is supported by the Director, Office of Science, Office of High Energy Physics of the U.S. Department of Energy under Contract No. DE-AC02-05CH1123, by the National Energy Research Scientific Computing Center, a DOE Office of Science User Facility under the same contract; and by the U.S. National Science Foundation, Division of Astronomical Sciences under Contract No. AST-0950945 to NOAO.

This publication makes use of data products from the Wide-field Infrared Survey Explorer, which is a joint project of the University of California, Los Angeles, and the Jet Propulsion Laboratory/California Institute of Technology, and NEOWISE, which is a project of the Jet Propulsion Laboratory/California Institute of Technology. WISE and NEOWISE are funded by the National Aeronautics and Space Administration.
%\end{acknowledgments}


\vspace{5mm}
\facilities{VLA, Arecibo}


\software{CASA~\citep{2022PASP..134k4501C}, 
          CARTA~\citep{2021zndo...3377984C},
          MIRIAD~\citep{1995ASPC...77..433S}
          Astropy~\citep{2013A&A...558A..33A,2018AJ....156..123A,2022ApJ...935..167A}}



%\appendix

%\section{Appendix information}


\bibliography{VirgoDarkDwarfs}{}
\bibliographystyle{aasjournal}



\end{document}